\documentclass{article}
\newtheorem{theorem}{Theorem}

\newtheorem{lemma}[theorem]{Lemma}
\newtheorem{corollary}[theorem]{Corollary}

\begin{document}


\title{
Generalized Landau-Pollak Uncertainty Relation
}

\author{Takayuki Miyadera$\ ^*$
and Hideki Imai$\ ^{*,\dagger}$
%
\\
$\ ^*$
Research Center for Information Security (RCIS), \\
National Institute of Advanced Industrial \\ 
Science and Technology (AIST). \\
Daibiru building 1102,\\
Sotokanda, Chiyoda-ku, Tokyo, 101-0021, Japan.
\\
(e-mail: miyadera-takayuki@aist.go.jp)
\\
$\ ^{\dagger}$
Graduate School of Science and Engineering,
\\
Chuo University. \\
1-13-27 Kasuga, Bunkyo-ku, Tokyo 112-8551, Japan .
}

\maketitle

\begin{abstract}
The Landau-Pollak uncertainty 
relation treats a pair of rank one projection valued measures
and imposes a restriction on their probability distributions.
It gives a nontrivial bound for summation of their 
maximum values. 
We give a generalization of this 
bound (weak version of the Landau-Pollak uncertainty 
relation). Our generalization
 covers a pair of positive operator valued measures. 
A nontrivial but slightly weak inequality that can treat 
an arbitrary number of positive operator valued measures is 
also presented.  
A possible application to the problem of 
separability criterion is also suggested.  
\end{abstract}

\section{Introduction}
The uncertainty relation represents one of the most fundamental 
aspects of the quantum theory. It is not only interesting 
by itself  
but also plays crucial roles in the various fields. 
The uncertainty relation
imposes a restriction on probability distributions of 
measurement outcomes 
with respect to two (or more) 
noncommutative observables.
In addition to the most famous Robertson-type uncertainty 
relation \cite{Robertson}, 
there are a several types of the relation, 
such as entropic type \cite{MU,Uffink,Deutsch,KP,arxiv} and 
Landau-Pollak type \cite{MU,Uffink}. 
The difference among these uncertainty relations 
lies on the quantities measuring the randomness (unsharpness)
of the probability distributions.
In the Landau-Pollak type, the randomness of a 
probability distribution $\{p_i\}$ is 
characterized by its maximum value
$\max_i p_i$. 
The Landau-Pollak uncertainty relation shows that 
the maximum values with respect to a pair of 
noncommutative observables must satisfy a nontrivial inequality. 
From the inequality, a weaker but simpler 
nontrivial inequality for the summation of 
the maximum values is obtained. 
According to the weak version of
 Landau-Pollak uncertainty relation,  
the probabilities 
of two noncommutative observables cannot have 
$1$ 
as their maximum values simultaneously. 
In spite of the simplicity of the theorem for 
a pair of observables, no nontrivial generalization to 
three or more observables has been known yet.
More precisely, the Landau-Pollak uncertainty relation has been able to 
treat only pairs of rank one projection valued measures so far. 
In this paper, we try to generalize the weak version of the 
relation to cover an 
arbitrary number of positive operator valued measures. 
That is, our generalization is two-fold. It 
enables us to treat general types of observables, and 
gets rid of the 
restriction
on 
the numbers of observables. 
This paper is organized as follows.
In the next section we give our main results. 
We first show a lemma that plays key roles
in our generalization. 
We give a theorem for a pair of general observables 
that enables us to derive the weak version of the  Landau-Pollak 
uncertainty relation. 
Another theorem for an arbitrary numbers of general 
observables is also presented. It relates the bound with 
pairwise noncommutativity. Although the theorem for 
an arbitrary number of general observables is 
not trivial, it is not strong enough to 
derive the original weak version of the 
Landau-Pollak uncertainty relation
for a pair of observables. 
In section \ref{sect:discussions},
we show a several examples and a possible application of 
our theorem to separability criterion in qudit system. 
\section{Main Results}
A POVM $A$ (on a discrete space) is a family of 
positive operators $\{A_i\}$ satisfying 
$\sum_{i} A_i ={\bf 1}$. 
Although, for simplicity, 
the measures are restricted to the discrete case,
it is straightforward to generalize our theorems to 
the general (continuous) measurable space. 
POVMs give the most general 
description of the observables. An important subclass
of the observables is a class of 
projection valued measures (PVMs). 
A PVM is a family of projection operators 
whose summation equals the identity operator. 
Suppose we have a POVM: $A=\{A_i\}$. 
For any state $\rho$, $\langle A_i \rangle_{\rho}
:=\mbox{tr}(\rho A_i)$ means the probability of 
getting an outcome $i$ when we measure the observable $A$ 
in the state $\rho$. 
\par
In its original version \cite{LP} 
interpreted by Maassen and Uffink\cite{MU,Uffink}, 
the Landau-Pollak 
uncertainty relation treats the simplest 
projection valued measures (PVMs) $P:=\{P_i\}$ and 
$Q=\{Q_j\}$, where $P_i$'s and $Q_j$'s are rank one 
projection operators as 
$P_i=\{|i\rangle\langle i|\}$ and 
$Q_j=\{|\tilde{j}\rangle \langle \tilde{j}|\}$.
They derived an inequality: 
\begin{eqnarray*}
\mbox{Arccos}\langle P_i \rangle_{\rho}
+\mbox{Arccos}\langle Q_j \rangle_{\rho} 
\geq \mbox{Arccos}|\langle i|\tilde{j}\rangle |.
\end{eqnarray*}
By maximizing $\langle P_i\rangle_{\rho}
+\langle Q_j\rangle_{\rho}$ under this bound,
one can obtain 
a weaker but simpler inequality:
\begin{eqnarray}
\langle P_i \rangle_{\rho}
+\langle Q_j \rangle_{\rho} 
\leq 1+|\langle i|\tilde{j}\rangle |. \label{weakLP}
\end{eqnarray}
To relate it with generalized entropic quantities, 
Vicente and S\'anchez-Ruiz \cite{separability}
 introduced a quantity $M_{\infty}(P:\rho)
=\max_i \langle P_i\rangle_{\rho}$ and showed 
an inequality for rank one PVMs $P$ and $Q$, 
\begin{eqnarray*}
M_{\infty}(P:\rho)+M_{\infty}(Q:\rho)
\leq 1+ \max_{i,j} |\langle i|\tilde{j}\rangle|.
\end{eqnarray*}
In what follows, we give a generalization of the 
weak version of the Landau-Pollak uncertainty relation (\ref{weakLP}).
The following lemma plays essential roles in the generalization.
\begin{lemma}\label{keylemma}
Let $P_1,P_2,\ldots,P_m$ be projection operators on a Hilbert space ${\cal H}$,
and let $|\psi\rangle$ be some unit vector in ${\cal H}$
satisfying $P_j |\psi\rangle \neq 0$ for all $j$.
Let 
$\Lambda(|\psi\rangle)$ denote the maximal eigenvalue of the 
positive semidefinite $m\times m$ matrix $G$ given by 
\begin{eqnarray*}
G_{ij}:=
\frac{\langle \psi|P_i P_j |\psi\rangle}
{\Vert P_i| \psi\rangle\Vert \cdot \Vert P_j |\psi\rangle\Vert}.
\end{eqnarray*}
Then 
\begin{eqnarray*}
\sum_{i=1}^m \langle \psi|P_i |\psi\rangle 
\leq \Lambda(|\psi\rangle )
\end{eqnarray*}
holds. 
\end{lemma}
{\bf Proof:}
\par
To show the positive semidefiniteness of the matrix $G$, we 
define a normalized vector $|\psi_i\rangle$ for each $i$
as 
\begin{eqnarray*}
|\psi_i\rangle :=\frac{P_i |\psi\rangle}{\Vert P_i |\psi\rangle\Vert }.
\end{eqnarray*}
Since the matrix $G$ can be written as 
$G_{ij} =\langle \psi_i|\psi_j
\rangle$, for all $x_1,x_2,\ldots,x_m \in {\bf C}$, 
$\sum_{i,j=1}^m \overline{x_i} G_{ij}x_j
=\Vert \sum_{i=1}^m x_i |\psi_i\rangle\Vert^2 \geq 0$ holds. 
The matrix $G$ is thus obviously positive semidefinite. 
Thanks to this semidefiniteness, 
we have for all $x_1,x_2,\ldots,x_m \in {\bf C}$, 
\begin{eqnarray*}
\sum_{i,j=1}^m \overline{x_i}G_{ij} x_j
\leq \Lambda(|\psi\rangle)\sum_{i=1}^m |x_i|^2.
\end{eqnarray*}
Hence 
\begin{eqnarray*}
\left(
\sum_{i=1}^m \langle \psi|
P_i|\psi\rangle \right)^2
&=&\langle \psi|\sum_{i=1}^m P_i|\psi\rangle^2 
\\
&\leq& \langle\psi|(\sum_{i=1}^m P_i)^2
|\psi\rangle
\\
&=& \sum_{i,j=1}^m 
\langle \psi|P_iP_j|\psi \rangle
\\
&=&
\sum_{i,j=1}^m \Vert P_i|\psi\rangle \Vert \cdot
G_{ij} \cdot \Vert P_j|\psi\rangle \Vert
\\
&\leq &\Lambda(|\psi\rangle)
\sum_{i=1}^m \Vert P_i|\psi\rangle\Vert^2
\\
&=&
\Lambda(|\psi\rangle)\sum_{i=1}^m \langle \psi|P_i|\psi\rangle
\end{eqnarray*}
holds, where the inequality in the second line follows from 
a relation, 
$|\langle \psi| A|\psi \rangle|^2 \leq \langle \psi|A^2|\psi\rangle$ 
that holds for an arbitrary self-adjoint operator $A=A^*$. 
\hfill Q.E.D.
\par
The estimation of the maximal eigenvalue $\Lambda(|\psi\rangle)$ 
of the above lemma gives a
generalization of the weak version of the 
Landau-Pollak uncertainty relation. 
In case $m=2$, the exact diagonalization of $G$ can be 
easily performed. Due to this fact, we can 
generalize the uncertainty relation
 for a pair of general observables.
To obtain the theorem for a pair of POVMs, one needs 
an extension technique to represent positive operators 
as projection operators in an enlarged space. To clarify 
the argument, we first begin with a generalization 
to a pair of PVMs that is
 obtained by 
a direct application of Lemma \ref{keylemma}, and next 
show a theorem for a pair of general observables. 
\begin{theorem}\label{th0}
Let $P_1$ and $P_2$ be projection operators 
on a Hilbert space ${\cal H}$. 
For any state $\rho$,
\begin{eqnarray*} 
\langle P_1\rangle_{\rho}
+\langle P_2\rangle_{\rho} 
\leq 1+\Vert P_1 P_2\Vert
\end{eqnarray*}
holds, where $\Vert \cdot \Vert$ represents the operator 
norm defined as $\Vert A\Vert:=\sup_{|\psi\rangle \neq 0}
\frac{\Vert A|\psi\rangle\Vert}{\Vert |\psi\rangle\Vert}$.
\end{theorem}
{\bf Proof:}
\par
Since an arbitrary state can be decomposed into a mixture of pure states,
it suffices to prove the inequality only for the case that 
the state $\rho$ is pure and is 
written with a normalized vector 
$|\psi\rangle$ as $\rho=|\psi\rangle \langle \psi|$. 
If either $P_1|\psi\rangle$ or $P_2|\psi\rangle$ equals $0$, 
the statement is trivially true. Thus we may assume that 
both $P_1|\psi\rangle \neq 0$ and $P_2|\psi\rangle \neq 0$ hold. 
We apply lemma \ref{keylemma} with $m=2$. 
The $2\times 2$ matrix $G$ can be written in this case as,
\begin{eqnarray*}
G:=\left(
\begin{array}{ccc}
1& \langle \psi_1|\psi_2\rangle\\
\langle \psi_2|\psi_1\rangle & 1
\end{array}
\right)\ ,
\end{eqnarray*}
where $|\psi_k 
\rangle :=\frac{P_k|\psi\rangle}{\Vert P_k|\psi\rangle\Vert}$ 
for $k=1,2$. Its maximal eigenvalue is easily obtained as 
$\Lambda(|\psi\rangle)
=1+|\langle \psi_1|\psi_2\rangle|$.
It can be bounded as 
$\Lambda(|\psi\rangle)
=1+|\langle \psi|P_1 P_2|\psi\rangle|
\leq 1+\Vert P_1 P_2\Vert$.
It ends the proof.
\hfill Q.E.D.  
\par
To accomplish 
a generalization of the weak version of the Landau-Pollak 
uncertainty relation to a pair of positive operator valued measures,
we 
need an extension technique as discussed below. 
\begin{theorem}\label{th1}
Let $A$ and $B$ be positive operators 
that satisfy $A \leq {\bf 1}$ and $B \leq {\bf 1}$
on a Hilbert space ${\cal H}$.
For an arbitrary state $\rho$, 
\begin{eqnarray*}
\langle A\rangle_{\rho}+\langle B\rangle_{\rho}
\leq 1+\Vert  A^{1/2} B^{1/2}\Vert
\end{eqnarray*}
holds. 
\end{theorem}
{\bf Proof:}
\par
Also in this case it suffices to prove the inequality 
only for the case that the state $\rho$ is pure. We 
hence write $\rho$ as $\rho=|\Omega\rangle \langle \Omega|$
with a unit vector $|\Omega\rangle$. 
If either $A|\Omega\rangle$ or $B|\Omega\rangle$ is vanishing, 
the statement is trivially true. We thus may assume that 
neither of them are vanishing. 
An extension of the Hilbert space enables us to 
represent the positive operators $A$ 
and $B$ as projection operators as follows. 
Let us consider an enlarged Hilbert space ${\cal K}:=
{\cal H}\oplus {\cal H}\oplus {\cal H}$ and operators $P_1$ and 
$P_2$ as, 
\begin{eqnarray*}
P_1 &:=&\left(
\begin{array}{ccc}
A& \sqrt{A({\bf 1}-A)}& 0 \\
\sqrt{A({\bf 1}-A)}& {\bf 1}-A & 0\\
0& 0& 0
\end{array}
\right),
\\
P_2&:=&
\left(
\begin{array}{ccc}
B& 0& \sqrt{B({\bf 1}-B)}\\
0&0&0\\
\sqrt{B({\bf 1}-B)}&0 &{\bf 1}-B\\
\end{array}
\right),
\end{eqnarray*}
where we have used the commutativity between $A$ and 
${\bf 1}-A$ (or, $B$ and ${\bf 1}-B$) to guarantee the 
well-definedness of $\sqrt{A({\bf 1}-A)}$ (or, $\sqrt{B({\bf 1}-B)}$). 
It can be easily checked that $P_1$ and $P_2$ are 
indeed projection operators in the enlarged space ${\cal K}$.  
We define a unit vector $|\psi\rangle:=|\Omega\rangle
 \oplus 0 \oplus 0$ and apply lemma \ref{keylemma} to them. 
 Since $\langle \Omega|A|\Omega\rangle=\langle \psi|P_1|\psi\rangle$ 
 and $\langle \Omega|B|\Omega\rangle=\langle \psi|P_2|\psi\rangle$ hold,
 we obtain, 
 \begin{eqnarray*}
 \langle \Omega|A|\Omega\rangle
 +\langle \Omega|B|\Omega\rangle
 \leq 
 1+|\langle \psi_1|\psi_2\rangle|, 
 \end{eqnarray*}
 where $|\psi_k\rangle:=\frac{P_k|\psi\rangle}{\Vert P_k|\psi\rangle\Vert}$ 
 (for $k=1,2$) is a unit vector in ${\cal K}$. 
 By using $P_1|\psi\rangle =
 A|\Omega\rangle \oplus \sqrt{A({\bf 1}-A)}|\Omega\rangle
 \oplus 0$
 and $P_2|\psi\rangle=B|\Omega\rangle\oplus 0 \oplus 
 \sqrt{B({\bf 1}-B)}|\Omega\rangle$, 
 we can calculate $|\langle \psi_1|\psi_2\rangle|$ as,
 \begin{eqnarray*}
 |\langle \psi_1|\psi_2 \rangle|
 &=&\frac{|\langle \Omega|AB|\Omega\rangle|}{
 \Vert A^{1/2}|\Omega\rangle\Vert \cdot \Vert
 B^{1/2}|\Omega\rangle\Vert}.
 \end{eqnarray*}
 Defining unit vectors $|\phi_1\rangle=\frac{A^{1/2}|\Omega\rangle}
 {\Vert A^{1/2}|\Omega\rangle\Vert}$ and 
 $|\phi_2\rangle=\frac{B^{1/2}|\Omega\rangle}{\Vert 
 B^{1/2}|\Omega\rangle\Vert}$, we obtain, 
 \begin{eqnarray*}
 \frac{|\langle \Omega|AB|\Omega\rangle|}{
 \Vert A^{1/2}|\Omega\rangle\Vert \cdot \Vert B^{1/2}|\Omega\rangle\Vert}
 =|\langle \phi_1|A^{1/2}B^{1/2}|\phi_2\rangle|
 \leq \Vert A^{1/2}B^{1/2}\Vert.
 \end{eqnarray*}
 It ends the proof. 
 \hfill Q.E.D.
 \par
 Note that we did not use theorem \ref{th0} to 
 prove theorem \ref{th1}. Theorem \ref{th0} 
actually is regarded as a corollary of theorem \ref{th1}.
Although the following statement is obvious,
it is mentioned
for clarifying the relation between 
our theorem and the original version (\ref{weakLP}).
 \begin{corollary}\label{cor1}
Let $A=\{A_i\}$ and $B=\{B_j\}$ be POVMs on a Hilbert space
${\cal H}$. 
For an arbitrary state $\rho$ and $i,j$, 
\begin{eqnarray*}
\langle A_i\rangle_{\rho}
+\langle B_j\rangle_{\rho}
\leq 1+\Vert A_i^{1/2} B_j^{1/2}\Vert
\end{eqnarray*} 
holds.
 \end{corollary}
 \par
 Application of 
 rank one PVMs $A:=\{P_i\}=\{|i\rangle\langle i |\}$ and 
 $B=\{Q_j\}=\{|\overline{j}\rangle \langle \overline{j}|\}$ 
 to the above corollary immediately recovers 
 the weak version of Landau-Pollak uncertainty relation (\ref{weakLP}).
 \par
 The lemma \ref{keylemma} enables us to 
 obtain a nontrivial bound 
 also for an arbitrary number of POVMs.
 Although it is hard to obtain 
 eigenvalues of the matrix $G$ for a general $m$, 
 we can estimate the maximal value in terms of 
 pairwise noncommutativity such as 
 $\Vert P_i P_j\Vert$ ($i\neq j$).
We obtain the following theorem.
\begin{theorem}\label{final}
Let us consider a Hilbert space ${\cal H}$ and 
a family of $m$ positive operators $\{A_j\}_{j=1}^m$ satisfying
$A_j \leq {\bf 1}$. For any state $\rho$,
\begin{eqnarray}
\sum_{i=1}^m 
\langle A_i\rangle_{\rho}
\leq 1+\left(
\sum_{i \neq j}\Vert A_i^{1/2}A_j^{1/2} \Vert^2
\right)^{1/2},
\label{ineqfinal}
\end{eqnarray}
holds. 
\end{theorem} 
{\bf Proof:}
\par
It suffices to prove only the case that the state $\rho$ is pure 
and is written as $\rho=|\Omega\rangle \langle \Omega|$ 
with a unit vector $|\Omega\rangle \in {\cal H}$.
We may assume that $A_j|\Omega\rangle \neq 0$ holds for all $j$.
In fact, $\langle A_i\rangle$ 
is zero for $i$ with $A_i|\Omega\rangle =0$ 
and does not contribute 
to the left hand side of the inequalilty (\ref{ineqfinal}). 
To make use of Lemma \ref{keylemma} also in this case, we 
first employ the extension technique as in the 
previous theorem. 
Let us consider an enlarged Hilbert space 
${\cal K}= {\cal H}\oplus {\cal H} \oplus \cdots 
\oplus {\cal H}$ ($m+1$ times). 
An operator $R$ on ${\cal K}$ can be represented 
in a matrix form as, 
\begin{eqnarray*}
R=\left(
\begin{array}{ccccc}
R_{00}& R_{01}& R_{02}& \cdots& R_{0m}\\
R_{10}& R_{11}& R_{12}& \cdots& R_{1m}\\
R_{20}& \cdot & \cdot& \cdots& \cdot\\
\cdot&\cdot& \cdot& \cdots&\cdot\\
R_{m0}&\cdot & \cdot& \cdots& R_{mm}
\end{array}
\right).
\end{eqnarray*} 
That is, $R$ acts on a vector $|\Phi\rangle
=|\Phi_0\rangle\oplus |\Phi_1\rangle 
\oplus \cdots \oplus |\Phi_m\rangle$
as, 
\begin{eqnarray*}
R|\Phi\rangle
=\sum_{j_0 =0}^m R_{0j_0}|\Phi_{j_0}\rangle 
\oplus \sum_{j_1= 0}^m R_{1 j_1} |\Phi_{j_1}\rangle
\oplus \cdots \oplus \sum_{j_m =0}^m R_{m j_m}
|\Phi_{j_m}\rangle.
\end{eqnarray*}
Let us define operators $P_k$ for $k=1,\ldots m$ 
by their matrix elements as,
\begin{eqnarray*}
(P_k)_{ij}=\left\{ 
\begin{array}{cc}
A_k & (i=j=0)\\
\sqrt{A_k ({\bf 1}-A_k)} & (i=k, j=0)\\
\sqrt{A_k ({\bf 1}-A_k)} & (i=0, j=k)\\
0& (\mbox{otherwise}).
\end{array}
\right.
\end{eqnarray*}
It is easy to check that they are projection operators in ${\cal K}$. 
For a unit vector $|\psi\rangle:=|\Omega\rangle
\oplus 0 \oplus 0 \oplus \cdots \oplus 0$, 
$\langle \psi|P_k|\psi\rangle
=\langle \Omega|A_k|\Omega\rangle$ holds,
hence we can apply Lemma \ref{keylemma} to obtain a bound 
for $\sum_k \langle \Omega|A_k|\Omega\rangle$.
Definition of unit vectors $|\psi_k \rangle := 
\frac{P_k|\psi\rangle}{
\Vert P_k|\psi\rangle\Vert}$ ($k=1,2,\ldots,m)$ allows 
an expression, $G_{ij}=\langle \psi_i|\psi_j\rangle$. 
Since each eigenvalue $\lambda_t$ $(t=1,2,\ldots,m)$
can be obtained by the diagonalization of 
this matrix, there exists a unitary matrix $u$ 
satisfying for each $t$ ,
\begin{eqnarray*}
\lambda_t&=&\sum_{ij}u_{ti}^* \langle \psi_i|\psi_j \rangle
u_{tj}
\\
&=&
\sum_i u_{ti}^*u_{ti} + 
\sum_{i\neq j} u_{ti}^* u_{tj} \langle \psi_i|\psi_j\rangle
,
\end{eqnarray*}
whose right hand side can be bounded by the
unitarity of $u$ and the Cauchy-Schwarz inequality as,
\begin{eqnarray*}
\lambda_t
&\leq&
1+ \left(
\sum_{i\neq j} |u_{ti}^* u_{tj}|^2\right)^{1/2}
\left(
\sum_{i\neq j}
|\langle \psi_i |\psi_j \rangle|^2
\right)^{1/2}
\\
&\leq&
1+ \left(
\sum_{ij} |u_{ti}^* u_{tj}|^2\right)^{1/2}
\left(
\sum_{i\neq j}
|\langle \psi_i |\psi_j \rangle|^2
\right)^{1/2}
\\
&\leq &
1+ \left(
\sum_{i\neq j}
|\langle \psi_i |\psi_j \rangle|^2 \right)^{1/2}.
\end{eqnarray*}
Since the above inequality holds for all $t$,
$\Lambda(|\psi\rangle)\leq  1 +\left(
\sum_{i \neq j}
|\langle \psi_i |\psi_j \rangle|^2 \right)^{1/2}$
holds. 
As in the proof of theorem \ref{th1}, 
$|\langle \psi_i|\psi_j\rangle|$ can be bounded 
as,
\begin{eqnarray*}
|\langle \psi_i |\psi_j \rangle|
=\frac{|\langle \Omega|A_i A_j|\Omega\rangle|}
{\Vert A_i^{1/2}|\Omega\rangle
\Vert \cdot \Vert A_j^{1/2} |\Omega\rangle \Vert}
\leq \Vert A_i^{1/2} A_j^{1/2}\Vert.
\end{eqnarray*}
It ends the proof.
\hfill Q.E.D.
\par
As discussed in the next section, this bound is
not strong enough although it is nontrivial.
\par
The following corollary is obvious.
\begin{corollary}\label{corfinal}
Let $A^{(j)}=\{A^{(j)}_s\}_s$ be a POVM for each $j=1,2,\ldots,m$
on a Hilbert space ${\cal H}$. For an arbitrary state $\rho$ and 
arbitrary $s_1,s_2,\ldots,s_m$, 
\begin{eqnarray*}
\sum_{j=1}^m \langle A^{(j)}_{s_j}\rangle_{\rho}
\leq 1+ \left(
\sum_{i \neq j}
\Vert (A^{(i)}_{s_i})^{1/2}
(A^{(j)}_{s_j})^{1/2}
\Vert^2
\right)^{1/2}
\end{eqnarray*}
holds.
\end{corollary}
\section{Discussions}\label{sect:discussions}
In this section, we discuss a several examples and 
suggest a possible application to the problem of 
separability criterion. 
\par
First let us consider the relationship between 
theorem \ref{th1} (or corollary \ref{cor1}) 
and theorem \ref{final} (or corollary \ref{corfinal}).
Since in theorem \ref{final} the number of observables 
is arbitrary, one may put it as $m=2$. Then we obtain,
for positive operators $A$ and $B$, 
$\langle A\rangle_{\rho}+\langle B\rangle_{\rho}
\leq 1+\sqrt{2}\Vert A^{1/2} B^{1/2}\Vert$, which is 
slightly worse than theorem \ref{th1}.
\par 
Next let us consider the most trivial case that 
$\Vert A^{1/2}_i A^{1/2}_j\Vert =0$ holds for all $i\neq j$. 
For instance, if $A= \{A_i\}$ itself forms a PVM, 
the condition $\Vert A^{1/2}_i A^{1/2}_j\Vert =0$ 
for all $i\neq j$ follows.
It gives the trivial bound, $\sum_i \langle A_i\rangle_{\rho}
\leq 1$. 
Although also in the case $A=\{A_i\}$ forms a POVM 
this trivial bound $\sum_i \langle A_i \rangle_i 
\leq 1$ should hold, unfortunately 
our inequality is not strong enough to show it.
\par
As an example to show the non-triviality 
of our theorem, we consider a $D$-dimensional Hilbert space 
and its mutually unbiased basis (MUB).
A MUB \cite{Wootters} is a family of bases $\{{\cal B}^{(j)}\}$, 
where ${\cal B}^{(j)}=\{|1:j\rangle, |2:j\rangle,
\cdots,|D:j\rangle\}$ is an orthonormalized basis. 
They satisfy, for $i \neq j$,
\begin{eqnarray*}
|\langle s:j|t:i\rangle|=\frac{1}{\sqrt{D}}.
\end{eqnarray*}
It is a longstanding problem to 
ask how many bases are compatible with respect to this 
condition. It is known 
that its upper bound is $D+1$ for $D$-dimensional 
Hilbert space, and for $D=p^r$ ($p$ is a prime) this upper bound is 
actually 
attained \cite{Wootters,Ivanovic}. 
We assume that we treat a Hilbert space that 
attains this upper bound $D+1$. 
Let us take a vector $|s_i:i\rangle$ from each basis
${\cal B}^{(i)}$, and put 
$P_i:=|s_i:i\rangle\langle s_i:i|$. 
According to our theorem, 
we obtain,
\begin{eqnarray}
\sum_{i=1}^{D+1} \langle P_i \rangle_{\rho}
\leq 1+ \sqrt{D+1}. \label{MUB}
\end{eqnarray}
Let us compare it with the bound that is trivially 
obtained by combining the inequality 
for a pair of observables. 
According to theorem \ref{th1}, for each pair of $i$ and $j$ $(i \neq j)$, 
\begin{eqnarray*}
\langle P_i \rangle_{\rho}
 +\langle P_j\rangle_{\rho} \leq 1+\sqrt{ \frac{1}{D}}
\end{eqnarray*}
 holds. Trivially combining them into an inequality gives,
 \begin{eqnarray*}
 \sum_i \langle P_i \rangle_{\rho}
 \leq \frac{D+1}{2}+\frac{1}{2}\left( \sqrt{D}+\frac{1}{\sqrt{D}}\right).
 \end{eqnarray*}
One can 
easily see that 
our inequality (\ref{MUB}) gives better bound for $D\geq 3$, and 
the discrepancy between our upper bound and 
the bound obtained by the trivial combination becomes 
larger for larger dimension $D$. 
An analogous discussion has been done for the entropic uncertainty relations,
and the nontrivial bound for MUB has been 
obtained\cite{IvanovicEntropic,Sanchez-RuizEntropic}.
\par
Finally we make a comment on a possible application of 
our generalized Landau-Pollak uncertainty relation
in the quantum information. 
The problem of finding a criterion to 
distinguish between separable states and  entangled states 
is one of the 
important problems. Recently, criteria based upon 
the various uncertainty relations 
\cite{HofmannTakeuchi,Guhne,Giovannetti,GandL,separability} 
have been proposed. 
Among them, Vicente and S\'anchez-Ruiz \cite{separability} employed 
the Landau-Pollak uncertainty relation. While they examined the 
strength of their criterion for bipartite qubit system, they proposed 
a criterion for general bipartite ($D$-dimensional) qudit system 
((57) in \cite{separability}). As they suggested there, a generalization 
of Landau-Pollak uncertainty relation to arbitrary numbers of observables 
enables us to propose another criterion for $D=p^r$ ($p$: prime)
as follows. 
 Let us consider a MUB $\{{\cal B}^{(j)}\}$ on ${\cal H}_A$, 
where ${\cal B}^{(j)}=\{|1:j\rangle, |2:j\rangle,
\cdots,|D:j\rangle\}$ is an orthonormalized basis 
satisfying, for $i \neq j$,
$|\langle s:j|t:i\rangle|=\frac{1}{\sqrt{D}}$.
One can define a self-adjoint operator $A^{(j)}$ as 
$A^{(j)}:=\sum_s f(s)|s:j\rangle \langle s:j|$ for 
a suitable function $f$. Its copy on ${\cal H}_B$ is also defined 
and written as, $B^{(j)}$. 
We write the maximum probability of the outcome with respect to 
an observable $A$ and a state $\rho$ as $M_{\infty}(A:\rho)$. 
For separable states, 
they satisfy,
\begin{eqnarray*}
\sum_i M_{\infty}(A^{(i)}\otimes B^{(i)}:\rho^{\mbox{sep}})
\leq 1+\sqrt{D+1}.
\end{eqnarray*} 
It is a natural extension of the separability criterion 
suggested in \cite{separability}. The detailed 
formulation and analysis will be appeared 
elsewhere.
\par
In conclusion, we showed a generalization of the weak version of the 
Landau-Pollak 
uncertainty relation. Our generalized inequality gives a nontrivial bound 
for  
an arbitrary number of positive operator valued measures. 
A possible application to the problem of separability criterion 
was also suggested.
\section*{Acknowledgments}
The authors thank Dr. Hans Maassen and the other 
anonymous referee for their fruitful comments
which remarkably improved the arguments presented in this paper. 

\end{document}